\documentclass[fleqn,12pt,twoside]{article}
\usepackage{espcrc1}

\usepackage{graphicx}
\usepackage[figuresright]{rotating}

\newcommand{\lsim}{<\kern-12pt\lower5pt\hbox{$\displaystyle\sim$}}
\newcommand{\gsim}{>\kern-12pt\lower5pt\hbox{$\displaystyle\sim$}}

\newcommand{\AmS}{{\protect\the\textfont2
  A\kern-.1667em\lower.5ex\hbox{M}\kern-.125emS}}

\title{Baryon superfluidity and neutrino emissivity of neutron stars}

\author{T. Takatsuka\address{Faculty of Humanities and Social Sciences, Iwate University, Morioka 020-8550, Japan 
        }\thanks{E-mail: takatuka@iwate-u.ac.jp}
        and  
        R. Tamagaki\address{Kamitakano Maeda-cho 26-5, Kyoto 606-0097, Japan}\thanks{E-mail: tamagaki@yukawa.kyoto-u.ac.jp}}
               
\begin{document}

\maketitle

\begin{abstract}
For neutron stars with hyperon-mixed core, neutrino emissivity is studied under the equation of state, 
obtained by introducing three-body force universal for all baryons so as to assure the  
maximum mass compatible with the observation. By paying attention to the density dependence of the critical 
temperatures of the baryon superfluids, which reflect the nature of baryon-baryon interaction and control 
neutron star cooling, we show what neutrino emission processes are efficient in the regions with and without 
hyperon mixing and remark its implications related to neutron star cooling.
\footnote{The present paper will be published in \textit{Proc. the 8th Int. Conf. on Clustering 
Aspects of Nuclear Structure and Dynamics, Nov. 24-29, 2003, Nara} to appear in Nucl. Phys. A. }

\end{abstract}

\section{HYPERON MIXING, EQUATION OF STATE AND COMPOSITION}

Hyperon ($Y$) mixing at high density in the core of neutron stars (NSs) brings about dramatic softening of 
the equation of state (EOS), because a substantial part of the repulsive contributions 
of the nucleon sector, growing at high density, is replaced by the attractive contributions of the nucleon-hyperon 
($NY$) and  hyperon-hyperon ($YY$) interactions since admixed hyperons are at low density. Under such too soft EOS 
the resulting maximum mass of NSs becomes smaller than the observed one $M_{\rm obs}=1.44M_{\odot}$, 
incompatible with the observation.   
This is the unavoidable feature appearing in an almost model-independent way. To get rid of this difficulty 
Nishizaki, Yamamoto and Takatsuka \cite{NYT02} introduced the repulsive part of three-body force working 
\textit{universally} among all the baryons (equally in the $NN$ , $NY$ and $YY$ parts),  
following the repulsive part (TNR) of the three-nucleon interaction (TNI) given by Pandharipande and collaborators 
who have shown the importance of three-body force in the nucleon system.\cite{LP81,HP00}  

\begin{figure}[htb]
\begin{minipage}[t]{75mm}
\includegraphics[width=0.95\linewidth]{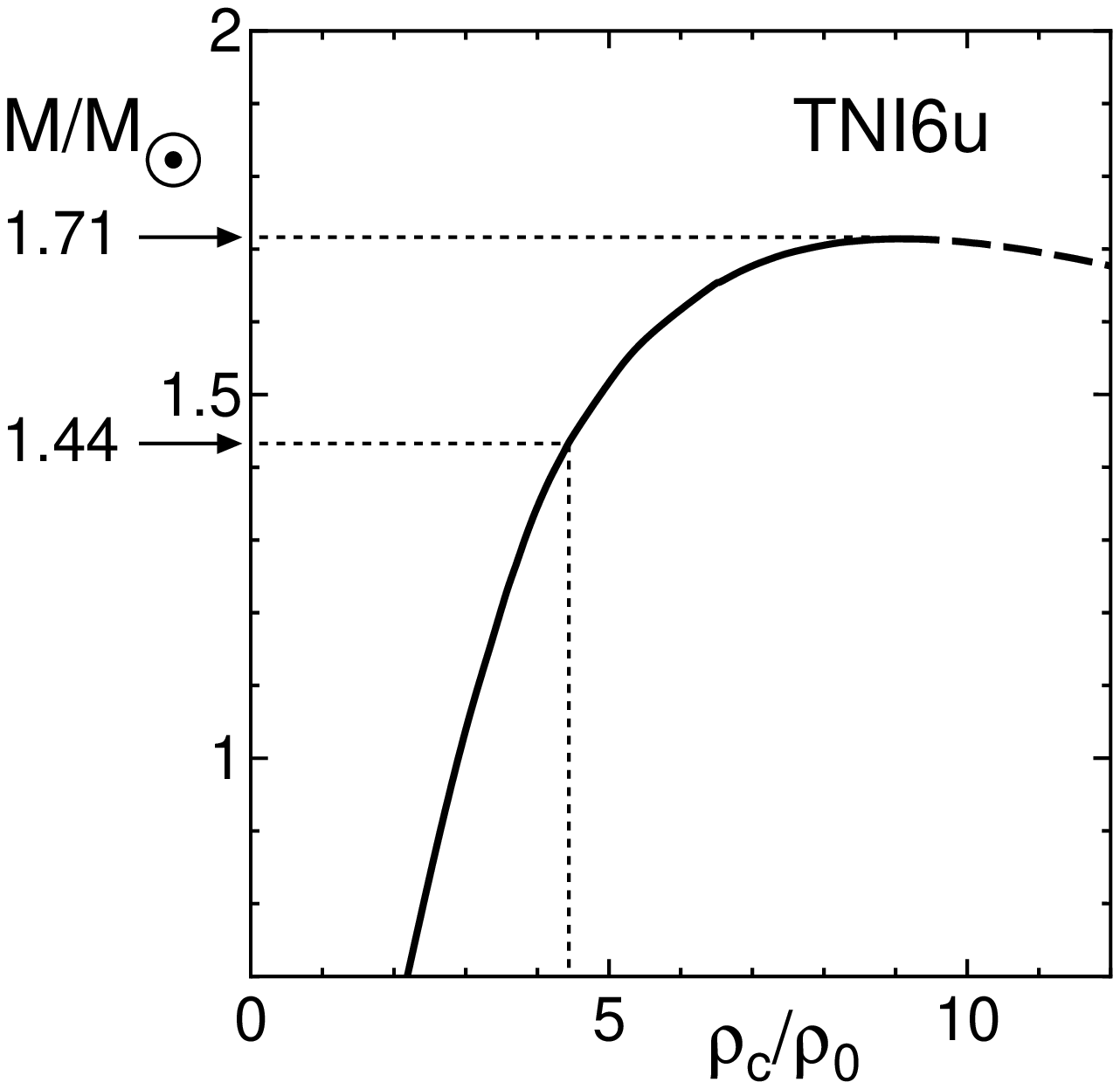}
\caption{Neutron star mass $M$ as a function of central density $\rho_{c}$ for the TNI6u-EOS.}
\label{fig:cdenNSM}
\end{minipage}
\hspace{\fill}
\begin{minipage}[t]{75mm}
\includegraphics[width=0.95\linewidth]{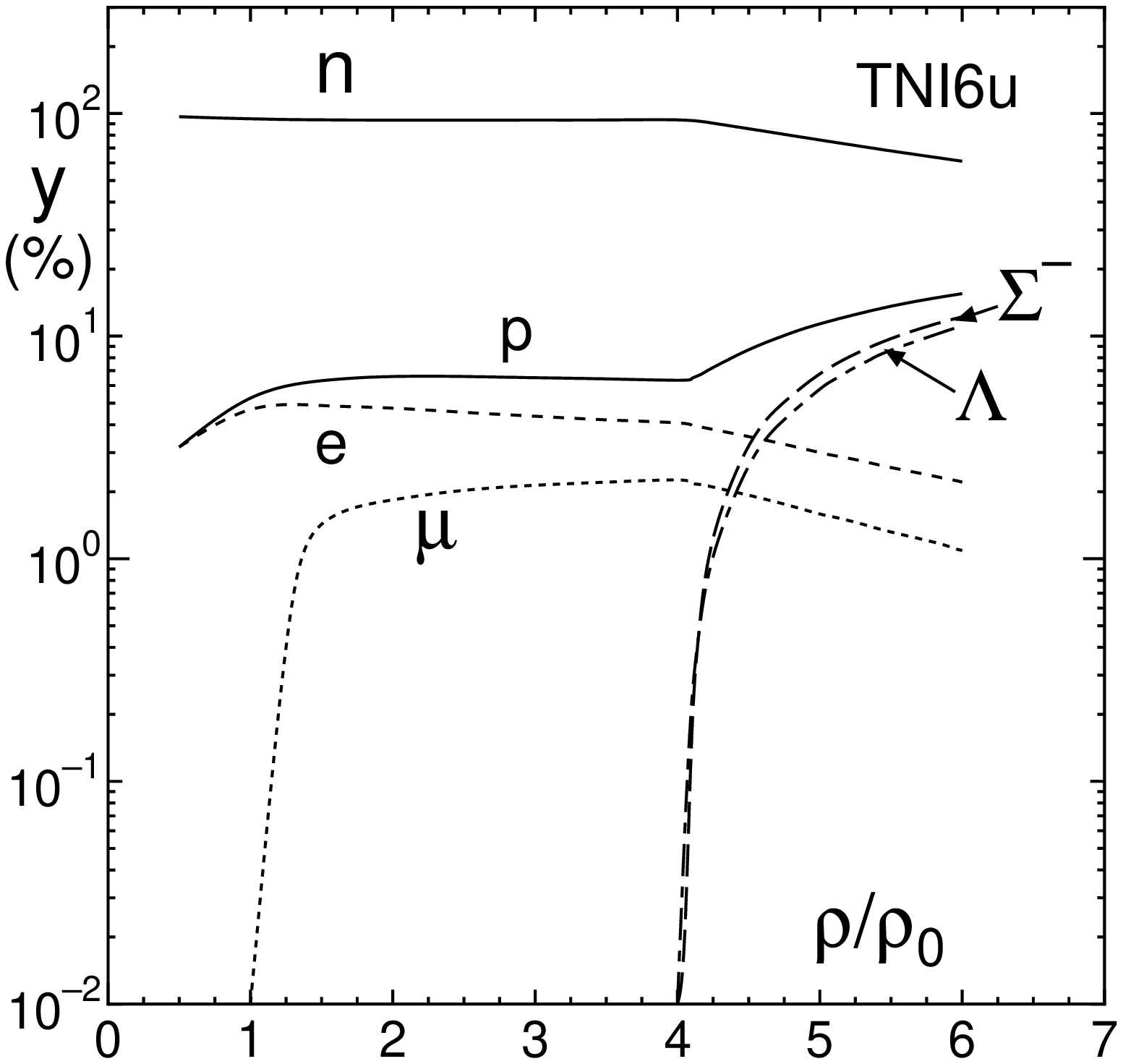}
\caption{Composition of neutron star matter;
fractions $y$ of constituent particles.}
\label{fig:Fractions}
\end{minipage}
\end{figure}

In the present discussion, we adopt the EOS denoted by TNI6u. The details are reported by Nishizaki 
in this conference (PB-16). Here its aspects directly related to the present study are remarked. 
The maximum value of NS mass ($M$) obtained for this EOS is $M_{\rm max}=1.71M_{\odot}$, as shown 
in Fig.\ref{fig:cdenNSM}. If we use the TNR only for the nucleon part, 
$M_{\rm max}\simeq 1.1M_{\odot}$ much less than $M_{\rm obs}$. Typical numbers 
of interest are the central densities $\rho_{c}$ corresponding to $M\sim M_{\rm obs}$. Fig.\ref{fig:cdenNSM} shows 
that $\rho_{c}\lsim 4\rho_{0}$ for $M\lsim 1.4M_{\odot}$, $\rho_{c}\simeq 4.5\rho_{0}$ for 
$M\simeq 1.44M_{\odot}$ and $\rho_{c}\gsim 5\rho_{0}$ for $M\gsim 1.5M_{\odot}$, $\rho_{0}$ being the nuclear density.

Composition of NS matter obtained under this EOS are shown by the fractions $y$ of the constituent particles 
in Fig. \ref{fig:Fractions}. Notable aspects are as follows:  Mixings of $\Lambda$ and $\Sigma^{-}$ set in 
at the almost same density $\rho\simeq 4\rho_{0}$ near the central density for the NS with $M\simeq 1.4M_{\odot}$. 
The fractions $y_{\Lambda}$ and $y_{\Sigma^{-}}$ increase abruptly to about 10 \% as 
$\rho \rightarrow 6\rho_{0}$ with a similar magnitude. 
Due to the $\Sigma^{-}$ mixing the proton fraction $y_{p}$ increases, but $y_{p}$ is 
still low so that the nucleon direct URCA process does not work until about $6.5\rho_{0}$.

\section{CRITICAL TEMPERATURES OF BARYON SUPERFLUIDS}

Generally energy gaps of the baryon superfliud (SF) are sensitive to the pairing interaction and the baryon effective mass. 
As the pairing interaction we select the $NN$ potentials reproducing the scattering phase shifts in the elastic 
region, AV18\cite{AV18} and OPEG\cite{OPEG}, and  the $YY$ potential 
 compatible with the available data on hypernuclei, ND-soft.\cite{NDsoft}  Concerning the effective masses 
we take the values determined by the EOS of TNI6u, here shown by the effective mass parameter $m_{i}^{*}$.\cite{NYT02}
 
For $p$, $\Lambda$ and $\Sigma^{-}$, because of their small fractions, the ${}^1 S_0$ SFs are realized for the 
fractional density $\lsim 0.5\rho_{0}$, with the upper limit determined mainly by the repulsive core effect.
 Critical temperatures $T_{c}^{(i)}$ are given by the energy gaps
$\Delta_{i}({}^1 S_{0})$ at zero temperature (taking the Boltzmann constant=1):
\begin{equation}
T_{c}^{(i)}\simeq 0.57\Delta_{i}({}^1 S_{0})\hspace{1cm}(i=p,\;\Lambda,\;\Sigma^{-}).
\end{equation}

For neutrons in the NS core ($\rho\gsim 0.7\rho_{0}$), the ${}^1 S_{0}$ SF disappears due to the repulsive core effect 
and the ${}^3 P_{2}$ SF is realized up to several $\rho_{0}$. The following two typical pairings with different 
angluar momentum components ($J=2,\;m_{J}$) are considered:
\begin{equation} 
T_{c}^{(n)} \simeq 0.60\Delta_{n}({}^3 P_{2},\;m_{J}=0),\;\; {\rm and}\;\; 
T_{c}^{(n)} \simeq 0.61\Delta_{n}({}^3 P_{2},\;m_{J}=\pm2),
\end{equation}
where $\Delta_{n}({}^3 P_{2},\;m_{J})$ is the energy gap at zero temperature. The critical temperature is the same, 
although the angular functions of the anisotropic energy gaps are different for the two cases. 
 
\begin{figure}[htb]
\begin{minipage}[t]{75mm}
\includegraphics[width=0.95\linewidth]{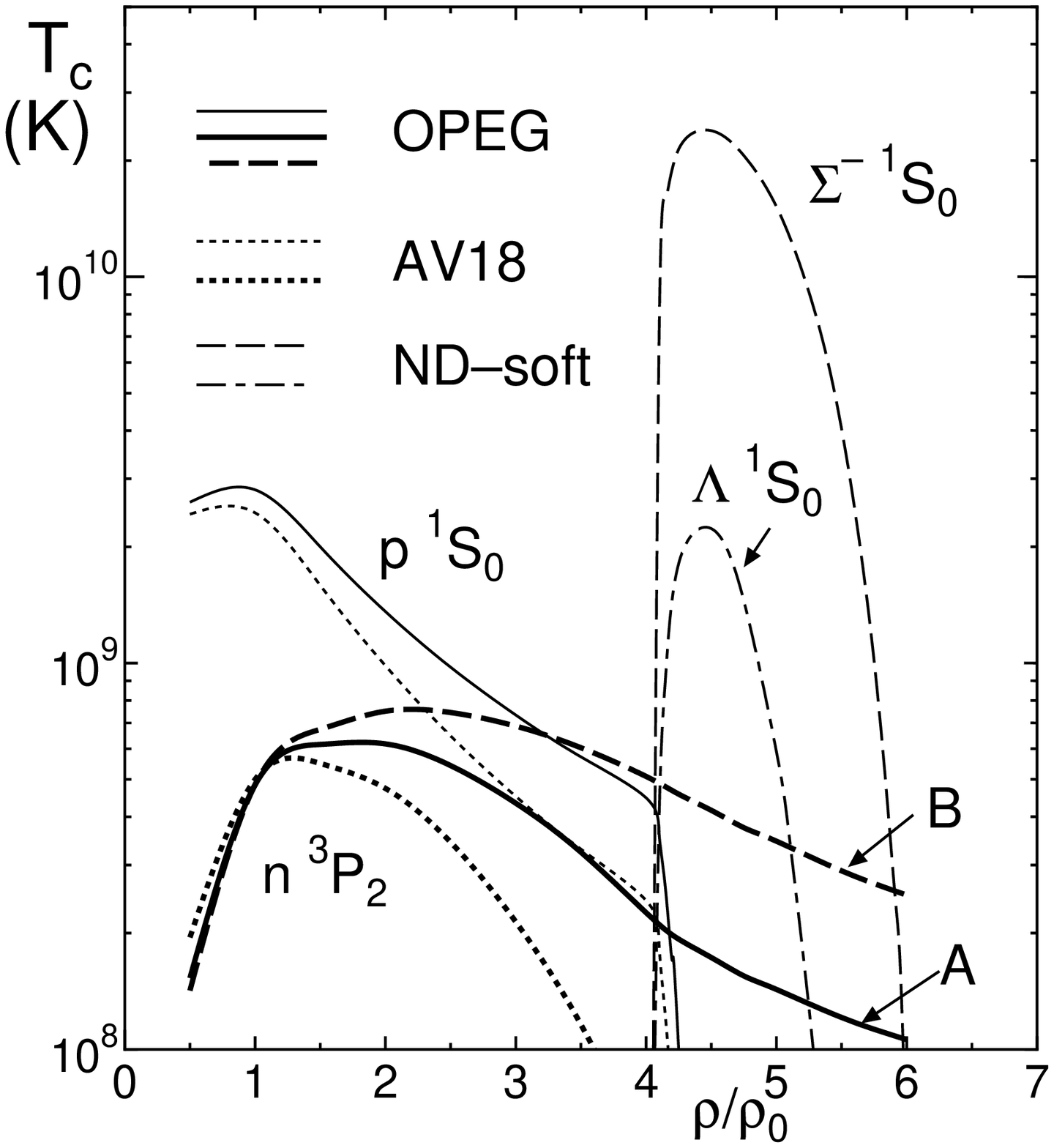}
\caption{Density dependence of critical temperatures $T_{c}$ of baryon superfluids. Variation seen between two curves 
for $p\; {}^1 S_{0}$ (OPEG-A and -B are the same in this state) and among three curves for $n {}^3 P_{2}$  mainly reflects 
the difference in the short-range repulsion.}
\label{fig:TcDdep}
\end{minipage}
\hspace{\fill}
\begin{minipage}[t]{75mm}
\includegraphics[width=0.8\linewidth]{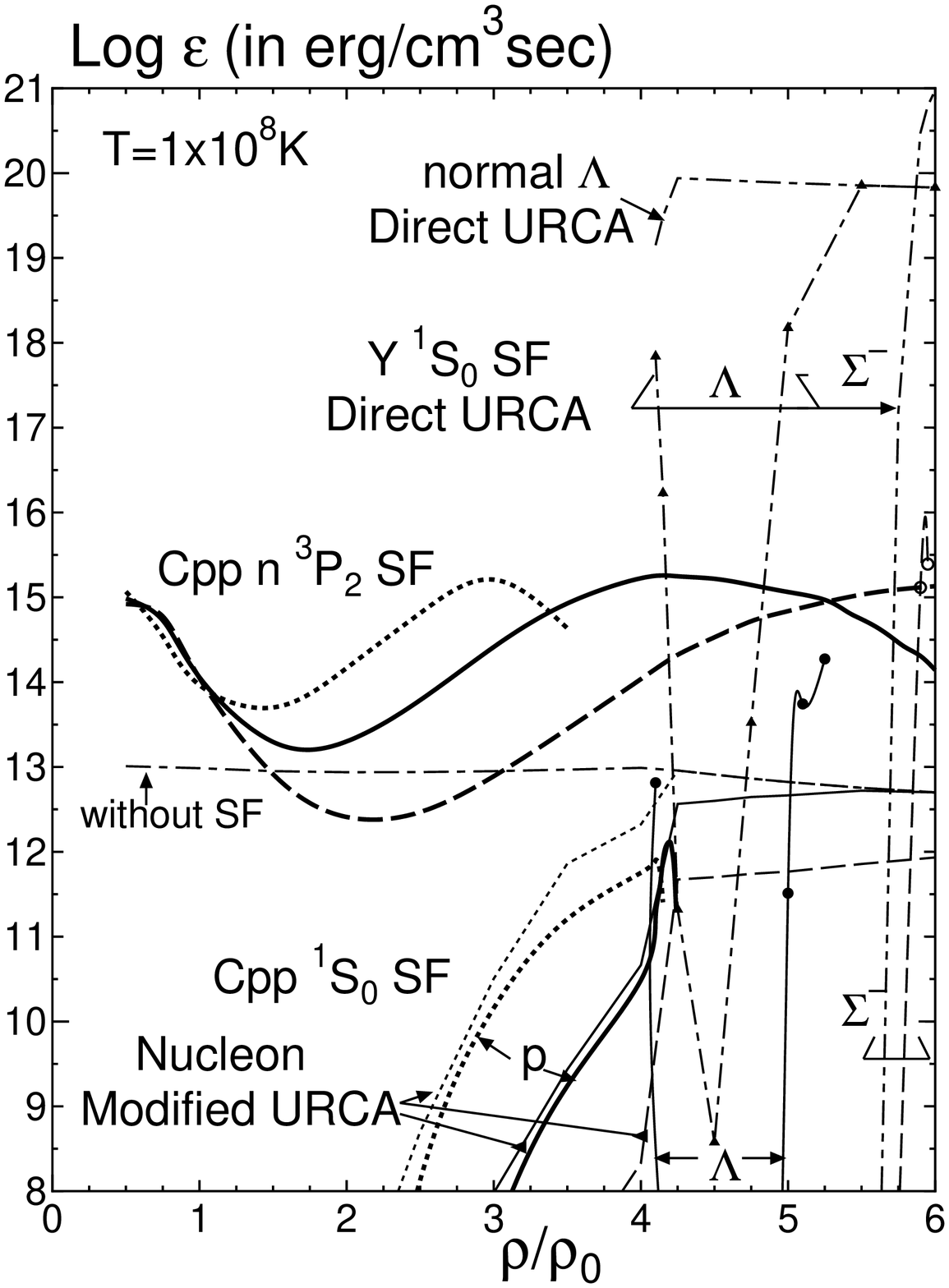}
\caption{Neutrino emissivities of three processes as functions of density. The dotted, solid and dashed curves are 
for AV18, OPEG-A and OPEG-B, respectively in the nucleon modified URCA (thin lines) and nucleon Cooper-pair 
(bold lines) processes. The dash-dot (dash-dot-dot) curves are for $\Lambda$ ($\Sigma^{-}$) direct URCA.} 
\label{fig:Emissibity}
\end{minipage}
\end{figure}

In Fig.\ref{fig:TcDdep}, the following points are remarked as to the $\rho$-dependence of $T_{c}^{(i)}$: 

(1) For the proton, $T_{c}^{(p)}$ calculated using the $NN\;{}^1 S_{0}$ potentials of AV18 and OPEG, gradually decreasing 
as $\rho$ goes high, drop sharply just beyond the threshold of $\Sigma^{-}\simeq 4\rho_{0}$, because the increase of $y_{p}$ 
(thus its Fermi momentum) makes the repulsive core effect stronger, in addition to the small effective mass 
as $m_{p}^{*}\simeq 0.6\rightarrow 0.5$ at $\rho\simeq(4\rightarrow 6)\rho_{0}$. 

(2) For $\Lambda$, $T_{c}^{(\Lambda)}$ calculated using the ND-soft $\Lambda\Lambda$ potential is moderately large 
$\sim 10^{9}{\rm K}\simeq 0.1$MeV because of the moderately large $m_{\Lambda}^{*}\simeq 0.8$, but the $\Lambda$ SF 
exists only in the limited region $\rho\simeq (4.0-5.3)\rho_{0}$. 

(3) For $\Sigma^{-}$, $T_{c}^{(\Sigma^{-})}$ is very large ($\sim 10^{9}-10^{10}$K) due to the large 
$m^{*}_{\Sigma^{-}}\simeq (1.3\rightarrow 1.0)$ as $\rho\simeq (4\rightarrow 6)\rho_{0}$. 

(4) For the neutron ${}^3 P_{2}$ SF, $T_{c}^{(n)}$ calculated using AV18 and OPEG-A \& -B are of moderate magnitude, 
showing the peak values $\simeq (6-8)\times 10^{8}$K and decreasing gradually as $\rho$ increases. A variety shown 
by three curves means that the stronger the short-range repulsion, the smaller the energy gaps. We regard 
this extent as the reliable allowance in $T_{c}^{(n)}$ of the core neutrons that is acceptable from the viewpoint 
of nuclear theory.

\section{NEUTRINO EMISSIVITIES OF EFFICIENT $\nu$-EMISSION PROCESSES FOR HYPERON-MIXED NEUTRON STARS}

Here we discuss the calculated results of the emissivities ${\cal E}$ (in units of erg/cm${}^3\cdot$sec) due to three 
$\nu$-emission processes efficient in the NS core. They depend strongly on the density and internal 
temperature $T$.\cite{Tsuruta98,YKGH01} 
${\cal E}$ is written as a product of the factor containing $T^{n}$ 
with $n=6-8$ characteristic of each process and the factor representing the SF effect sensitive to the ratio $\Delta_{i}/T$,  
where $\Delta_{i}$ is the $T$-dependent energy gap.  As an example, we show Log${\cal E}$ in Fig. \ref{fig:Emissibity} for 
the typical internal temperature $T=1\times 10^{8}$K.

(1) Direct Urca (DURCA) process is described as   
$B_{1}\rightarrow B_{2}+\ell^{-}+\bar{\nu}_{\ell},\;\;(\ell=e,\mu)$ and its inverse.  In the baryon composition of  
TNI6u-EOS, the nucleon DURCA does not occur and the hyperon DURCA is possible only for the 
$\Lambda\rightarrow p$ and $\Sigma^{-}\rightarrow \Lambda$ decays. In the density region with  $T_{c}^{(\Lambda)}\gsim 5T$, 
the hyperon ${\cal E}_{\rm DURCA}$ is strongly suppressed by the energy gaps of $\Lambda$ and $\Sigma^{-}$. 
If the $\Lambda$ matter is normal, the enormous ${\cal E}_{\rm DURCA}\sim 10^{20}T_{8}^{6}$ comes about, 
where $T_{8}\equiv T/10^{8}$K.

(2) Nucleon modified URCA process (MURCA), playing important roles in the \textit{standard} scenario 
of NS cooling, is described as $n+B\rightarrow p+B+\ell^{-}+\bar{\nu}_{\ell},\;(\ell=e,\; \mu)$ and its inverse, 
where $B$ means a by-stander baryon. Here we treat the case of $B=n,\; p$. Without the SF suppression, 
${\cal E}_{\rm MURCA} \sim 10^{13}$ at $T_{8}=1$ increases by the factor $(T_{8})^{8}$. The nucleon SFs strongly 
suppress the emissibity at $\rho\lsim 3\rho_{0}$ where $T_{c}^{(p)}$ and $T_{c}^{(n)}$ are larger than 
about $5\times 10^{8}$K. Thus the nucleon MURCA brings about the weak NS cooling. 

(3) Cooper-pair process (Cpp) caused by the neutral current of the weak interaction is described as the process 
of $\nu\bar{\nu}$ emission when  two excited SF quasi-particles recombine into the Cooper-pair in the BCS 
state.\cite{FRS76,YKL99} This works only in the presence of SF of some baryon ($T_{c}^{(i)}>T$). 
${\cal E}_{Cpp}$ with $n=7$ is considerably large at $T<T_{c}^{(i)} \lsim 5T$ but, for $T_{c}^{(i)}\gg T$, is strongly 
suppressed by a factor $\sim {\rm exp}(-\Delta_{i} /T)$. The contributions due to the neutron ${}^3 P_{2}$ SF dominate 
in wide density region, over those from $p$, $\Lambda$ and $\Sigma^{-}$. 
However ${\cal E}_{Cpp}\sim 10^{1-2}{\cal E}_{\rm MURCA}$ is still moderate.

In comparison seen in Fig. \ref{fig:Emissibity} the following points are remarked: 
At $\rho\lsim 4\rho_{0}$, the neutron ${}^3 P_{2}$ SF Cooper-pair process dominates. At $\rho\gsim 4\rho_{0}$, 
the hyperon direct URCA process dominates if $T_{c}^{(\Lambda)}$ is not large.  
For $T_{c}^{(\Lambda)}\lsim T$, ${\cal E}_{\rm DURCA}(\Lambda\rightarrow p)$ is enormously efficient. 
The key point is the magnitude of $T_{c}^{(\Lambda)}$.
  
\section{CONCLUDING REMARKS RELATED TO NS COOLING}

The aspects of emissivities for NSs with and without hyperon-mixed core, which are mentioned in the previous section, 
have been taken into account in the recent calculations performed by Tsuruta and her collaborators\cite{Tsuruta03}  
concerning thermal evolution curves, which describe the change of surface photon luminosity ($L$) corresponding 
to surface temperature as a function of age of NSs ($t$). 
Comparison of the calculated results with the observational data leads to the following points to be remarked:
 
(1) The colder class NS data ($L^{\rm obs}\sim 10^{33\rightarrow 31.5}$ ergs/sec for $t=10^{3\rightarrow 5.5}$ years) 
are well reproduced by taking NSs of $M\simeq (1.5-1.6)M_{\odot}$ with the hyperon-mixed core providing the direct 
URCA $\nu$ emission, if suppressed properly  
by the $\Lambda$ SF with $T_{c}^{(\Lambda)}\sim 10^{9}$K. If we take a less attractive $\Lambda\Lambda$ interaction 
leading to $T_{c}^{(\Lambda)}\lsim$ internal temperature, we face serious contradiction with the observation. 

(2) The hotter class NS data ($L^{\rm obs} \sim 10^{34\rightarrow 32}$ ergs/sec for $t=10^{3\rightarrow 6}$ years) can be 
reproduced by taking NSs of $M\lsim 1.4M_{\odot}$ without the hyperon-mixed core, if the Cooper-pair cooling effect is 
qualitatively in balance with the effect due to vortex creep heating.  

(3) It is important to study the possibility that the small $\Lambda\Lambda$-bond energy in ${}^6 _{\Lambda\Lambda}{\rm He}$ 
suggested by ``NAGARA" event\cite{Takahashi01} is explained without reduction of the $\Lambda\Lambda$ ${}^1 S_{0}$ attraction, 
e.g. by introducing a repulsive $\Lambda\Lambda N$ three-body interaction. 

(4) If meson condensation (especially of pion) takes place, it gives rise to strong direct URCA coolingCand 
leads to the too rapid cooling, if the SF suppression is absent. Therefore, persistence of baryon SF in the meson-condensed 
phase is indispensable. 
   
Finally we make one additional remark. In the course of studying the Cooper-pair process, we found the mistaken statement 
widely spread that the neutral current of the weak interaction does not couple to $\Lambda$ and $\Sigma^{0}$, e.g. in 
\cite{Maxwell79,YKL99,YKGH01}. In the present study we have used the correct expression derived in a joint work 
with Tatsumi.\cite{TTT03}

The authors wish to thank S. Nishizaki, T. Tatsumi and Y. Yamamoto for their collaborations in the studies on 
hyperon-mixed neutron star matter. They also thank S. Tsuruta for cooperative discussions and sending us her Sydney 
Conference paper before publication. \\

\end{document}